\documentclass[aip,reprint]{revtex4-1}

\usepackage{amssymb}
\usepackage{amsfonts}
\usepackage{amsmath}
\usepackage{color}

\DeclareMathOperator{\sinc}{sinc}

\begin{document}

\title{Parasitic nonlinearities in photon pair generation via integrated spontaneous four-wave mixing: critical problem or distraction?}

\author{L. G. Helt$^{1,2}$, M. J. Steel$^{2}$, and J. E. Sipe}

\affiliation{$^1$Department of Physics and Institute for Optical Sciences, University of Toronto, 60 St. George St., Toronto, ON M5S 1A7, Canada \\
$^2$Centre for Ultrahigh bandwidth Devices for Optical Systems (CUDOS), MQ Photonics Research Centre, Department of Physics and Astronomy, Faculty of Science, Macquarie University, NSW 2109, Australia}

\begin{abstract}
We consider integrated photon pair sources based on spontaneous four-wave mixing and derive expressions for the pump powers at which various nonlinear processes become relevant for a variety of source materials and structures. These expressions serve as rules of thumb in identifying reasonable parameter regimes for the design of such sources. We demonstrate that if pump powers are kept low enough to suppress cross-phase modulation, multi-pair events as well as many other nonlinear effects are often also constrained to negligible levels.
\end{abstract}

\maketitle

\textit{Introduction. }Many photon pair sources rely on one of two nonlinear optical processes, namely spontaneous parametric downconversion (SPDC)~\cite{Tanzilli:2002, Suhara:2007, Nasr:2008, Horn:2012, Zhu:2012}, or spontaneous
four wave mixing (SFWM)~\cite{Liang:2007, Clemmen:2009, Xiong:2011, Xiong:2011:2, Azzini:2012, Engin:2012}. Photon pair generation via SPDC can be regarded as relatively ``clean''---it is rare that other nonlinear effects are present at the pump powers of interest, and designers of SPDC photon pair sources typically focus only on
managing phase-matching and multi-pair production. While SFWM, as a $\chi^{(3)}$ process, opens up a wider range of materials and platforms to designers of photon pair sources, there may also be a host of competing nonlinear effects to manage at the pump powers of interest. This can lead to both reduced efficiency and a more complicated quantum state of generated photons than is produced via SPDC.

Of course, in an actual photonic quantum information processing system the measured state of generated photons is influenced not only by the generation process, but also by photon processing (coupling losses, linear absorption and scattering losses, etc.), and photon detectors (inefficiencies, dark counts, after-pulsing, etc). However, processing losses are continually shrinking as fabrication techniques improve and circuits are placed on-chip, and the impact of detectors is well-understood in terms of simple statistical models~\cite{Bussieres:2008, Takesue:2010, Jennewein:2011}. So here we focus on complications arising within an integrated SFWM photon pair source itself. For example, multi-pair production from SPDC sources is known to be the main cause of errors in many photonic quantum-logic gates~\cite{Barbieri:2009}, and so we expect it to be one of the leading detrimental effects for SFWM sources as well. Additionally, in SFWM sources both two-photon absorption (TPA) and associated free-carrier absorption (FCA) can cause the loss of generated photons, self- and cross-phase modulation (SPM and XPM) can shift the optimal pump and collection frequencies, dispersion of the pump pulse can become important if short enough pulses are used, and spontaneous Raman scattering (SpRS) can create noise in the form of single photons. While some of these nonlinear processes have been theoretically studied in conjunction~\cite{Yin:2007, Brainis:2009, Silva:2013}, a realistic multi-mode, fully quantum mechanical treatment of any one, let alone all of them, remains a challenge. Managing these processes is also important experimentally as, in general, they reduce both the measured pair generation rate and coincidence to accidentals ratio (CAR) of a given source. There are a wide range of nonlinear materials for sources (see Table \ref{tab:materials} for some current and potential examples) and the source structure can be chosen from fibers, nanowires, rings or photonic crystals (see Table \ref{tab:sourceprops}). Combining this freedom with the diversity of pump inputs from continuous wave (CW) to pulses of a few ps, it is not \emph{a priori} obvious which processes should be the primary concern of a designer of SFWM photon pair sources.

As a simple example, a recent experiment~\cite{Xiong:2012} demonstrated the effect of TPA in limiting the count rate and suppressing the CAR of pairs generated in a silicon photonic crystal. The effect was absent in a comparable GaInP device~\cite{Clark:2013}, which has a much larger band gap. Note, however, that in the silicon device the onset of these TPA effects occurred at input pump powers \emph{above} those at which the CAR begins to fall off due to multi-pair generation, and thus in normal operation one would not encounter the additional TPA penalty. This leaves one wondering: was this a fortuitous result of the particular structure, or can this be expected in many if not all silicon pair-generation devices? It is this kind of question that interest us here.%
\begin{table}[tbp]
\caption{Nonlinear index $n_2$ and two-photon absorption coefficient $\beta_\text{TPA}$ of various systems used for pair generation at a wavelength of 1550 nm. Hyphens indicate negligible values.  In addition, for Si the free-carrier absorption coefficient $\sigma_\text{FCA}=1.45\times10^{-21}$ m$^{2}$ and the free-carrier lifetime $\tau _{c}=1$ ns.  These quantities are negligible for the other materials presented.}%
\begin{ruledtabular}
\begin{tabular}{ll c c c c}
\multicolumn{2}{l}{Parameter} & SiO$_{2}$~\cite{Agrawal:2007} & As$_{2}$S$_{3}$~\cite{Gai:2012,maddenpriv} & Diamond (D)~\cite{Mildren:2013} & Si~\cite{Yin:2007} \\
\hline
$n_{2}$ & [$10^{-20}$ m$^{2}$/W] & 3.2 & 290 & 5 & 600 \\
$\beta_\text{TPA}$ & [$10^{-12}$m/W] & - & $<$0.01 & - & 5 \\
\end{tabular}
\end{ruledtabular}
\label{tab:materials}
\end{table}%

In this Letter we develop inequalities that serve as rules of thumb for designing SFWM photon pair sources, thus clarifying the impacts of the various nonlinear processes mentioned above. We further demonstrate that if the pump power is kept low enough to suppress XPM, one is often automatically constraining the impact of many other nonlinear processes to negligible levels as well.

Our discussion is restricted in two ways. To limit the degrees of freedom we consider only single pump configurations, reserving dual-pump configurations for later work. We also intentionally do not include SpRS. As SpRS is linear in the pump power at low power, whereas pair generation is quadratic, the na\"{i}ve strategy to minimize its effect would be to \emph{increase} the pump power rather than keep it below some value. However, in the relevant amorphous materials where SpRS is a concern at all, the required power would be orders of magnitude in excess of the no-multi-pair limit. The only ways that we see to manage noise induced by Raman scattered photons are cooling~\cite{Liang:2007}, collecting photons within a low Raman gain window~\cite{Xiong:2011}, using dispersion engineering~\cite{Fulconis:2007,Gai:2012} or dual pump configurations~\cite{Garay-Palmett:2007} to produce large pump-signal frequency shifts, or simply avoiding amorphous materials altogether.

\textit{Pair Generation Probabilities. }Taking a channel waveguide and a microring resonator side-coupled to a channel waveguide as two structure examples, we can write the probability of pair production in the undepleted pump approximation and the limit of a low probability of pair production as~\cite{Helt:2012}%
\begin{widetext}
\begin{align}
N_{\text{pairs}}^{\rho}
\approx &\frac{\left( \gamma PL\right) ^{2}T^{2}}{8\pi ^{2}}\int_{0}^{\infty
}\text{d}\omega _{1}\int_{0}^{\infty }\text{d}\omega _{2}\frac{\omega
_{1}\omega _{2}}{\omega _{P}^{4}}\left\vert F\left( \omega _{1}\right)
\right\vert ^{2}\left\vert F\left( \omega _{2}\right) \right\vert
^{2}\left\vert \int_{0}^{\omega _{1}+\omega _{2}}\text{d}\omega \,\phi
_{P}\left( \omega \right) \phi _{P}\left( \omega _{1}+\omega _{2}-\omega
\right) \right.   \notag \\
&\times \left. \sqrt{\omega \left( \omega _{1}+\omega _{2}-\omega \right) }%
\sinc\left\{ \frac{\beta _{2}}{2}\left[ \left( \omega -\frac{\omega
_{1}+\omega _{2}}{2}\right) ^{2}-\left( \frac{\omega _{1}-\omega _{2}}{2}%
\right) ^{2}\right] L\right\} F\left( \omega \right) F\left( \omega
_{1}+\omega _{2}-\omega \right) \right\vert ^{2},  \label{N}
\end{align}%
\end{widetext}where $\rho =\text{C,R}$ denotes the channel or ring device. In this expression $\gamma $ is the usual nonlinear parameter that depends on the source material and structure \cite{Helt:2012, Agrawal:2007}, $T$ is the intensity FWHM of the pump pulse in time, $\phi _{P}\left( \omega \right) $ is the pump pulse wave form, $\beta _{2}=\left. d^{2}k/d\omega^{2}\right\vert _{\omega _{P}}$ is the group velocity dispersion (GVD) parameter, and $L$ is the length of the waveguide (where for a resonator of radius $R$, $L=2\pi R$). Here $P=\hbar \omega _{P}N_{\text{pump}}/T$ is the pump pulse energy divided by $T$ with $N_{\text{pump}}$ the average number of pump photons per pulse, and thus the average pump power is $P_\text{avg}=PfT$ where $f$ is the repetition rate or, equivalently, $fT$ is the duty cycle of the pump laser. Additionally, we have assumed that a Taylor expansion to second-order about the center frequency of the pump serves to specify $k(\omega)$ for all $\omega$ of interest. The field enhancement factor $F(\omega)$ is unity for the channel waveguide and is defined for a ring resonator below. While TPA, FCA, SPM, and XPM have been neglected, we note that \eqref{N} represents the maximum possible number of generated photon pairs per pump pulse. That is, for a process phase matched for a pump with a fixed center frequency, if any of these additional nonlinear effects are significant they must lead to a lower number of generated correlated photon pairs than in \eqref{N}. The inclusion of TPA and/or FCA will lead to a reduction of the pump power, and thus reduce the number of generated pairs, or directly reduce the number of generated pairs, whereas the inclusion of SPM and/or XPM will change the pump frequency for which SFWM is phase-matched, and thus reduce the number of generated pairs unless the pump frequency is altered.%

\begin{table}[tbp]
\caption{Source properties of various systems. All systems exist and have
been used for photon pair production except the pulsed pump ring. Hyphens indicate quantities not relevant for the particular source. Tildes indicate approximate values extracted from the indicated reference.}%
\begin{ruledtabular}
\begin{tabular}{ll c c c c}
\multicolumn{2}{l}{ ~} & Pulsed & CW & Pulsed & CW \\
\multicolumn{2}{l}{ }  & Fiber~\cite{Liang:2007} & Waveguide~\cite{Xiong:2011} & Ring & Ring~\cite{Azzini:2012} \\
\hline
$L$            &[m] & 300 & 0.071 & 80$\pi\times$10$^{-6}$ & 10$\pi\times$10$^{-6}$ \\
$A_\text{eff}$ &[$\mu$m$^{2}$] & $\sim$60 & 0.86 & 1 & 0.13 \\
$Q$            & ~ & - & - & 5000 & 7900 \\
$\lambda$      &[nm] & 1555.95 & 1549.315 & 1550 & 1558.5 \\
$T$            &[ps] & 5 & - & 0.1 & - \\
$\beta _{2}$   &[fs$^{2}$/mm] & $\sim$3 & - & 5 & - \\
$\gamma$       &[m$^{-1}$W$^{-1}$] & $\sim$0.0022 & 14 & 0.20 & 190 \\
\end{tabular}
\end{ruledtabular}
\label{tab:sourceprops}
\end{table}

\textit{Channel Waveguide Expressions. }To evaluate \eqref{N}, we consider a number of realistic limits. For a channel waveguide, the phase matching bandwidth $\Delta _{M}\approx 4\sqrt{a/\left( \left\vert \beta_{2}\right\vert L\right) }$, where $a\approx 1.8955$ is the positive root of $\sinc\left( x\right) =0.5$, is generally quite large (see Table \ref{tab:sourceprops}). The generation of frequency-uncorrelated photon pairs (i.e.\ near-unity Schmidt number \cite{Law:2004}) by setting the pump bandwidth equal to this phase matching bandwidth $\Delta _{M}$ is very difficult when $\Delta _{M}$ is large. As such, frequency-uncorrelated photon pairs are typically generated in channel waveguides using pulses with bandwidths $\Delta_{P}\ll \Delta _{M}$, where $\Delta _{P}\approx 4a/T$, followed by filtering~\cite{Liang:2007, Xiong:2011}. Furthermore, while the pump pulse duration $T$ has no influence on the factor in front of the integrals in \eqref{N}, long pulses are efficient at converting a fixed average number of pump photons $N_\text{pump}$ into generated photon pairs, as no pump energy falls outside the phase matching bandwidth in the integrand~\cite{Yang:2008}. In the long pulse limit, for a hard-edge filter of bandwidth $2\pi B$ with frequency detuning $\Omega $ from $\omega _{P}$ we find \cite{Helt:2012}%
\begin{equation}
N_{\text{pairs}}^{\text{C}}\left( \Omega \right) \approx \left( \gamma
PL\right) ^{2}TB\sinc^{2}\left( \beta _{2}\Omega ^{2}L/2\right) ,
\label{NC1}
\end{equation}%
or, integrating across the entire generation bandwidth \cite{Helt:2012,Brainis:2009}%
\begin{align}
N_{\text{pairs}}^{\text{C}}\approx & \left( \gamma PL\right) ^{2}\frac{2}{3}%
\sqrt{\frac{T^{2}}{2\pi \left\vert \beta _{2}\right\vert L}}  \notag \\
=& \left( \frac{L}{L_{\text{NL}}}\right) ^{2}\frac{2}{3}\sqrt{\frac{L_{\text{%
D}}}{2\pi L}},  \label{NC2}
\end{align}%
where we have introduced $L_{\text{NL}}\equiv \left( \gamma P\right) ^{-1}$ and $L_{\text{D}}\equiv T^{2}/\left\vert \beta _{2}\right\vert$ as the usual nonlinear length and dispersion length, respectively~\cite{Agrawal:2007}.

Equations~\eqref{NC1} and \eqref{NC2} allow us to make a number of conclusions about the impact of different nonlinear effects in channel waveguides. Since many quantum information processing tasks demand very low rates of multi-pair production, we begin by insisting that the source operates in the no-multi-pair limit, i.e.\ $N_{\text{pairs}}^\rho\ll 1$. If the filters are placed near enough to the pump that the squared sinc function in \eqref{NC1} is essentially unity, the no-multi-pair limit for filtered photon pair generation with a long pulse becomes $\left( L/L_{\text{NL}}\right) ^{2}TB\ll 1$, and, for fixed $T$ and $B$, requires $L\ll L_{\text{NL}}$. Similarly, if working in the long pulse limit and collecting over the entire generation bandwidth, the restriction $\Delta _{P}\ll \Delta_{M}$ is equivalent to the condition $L\ll L_{\text{D}}/a$, and so \eqref{NC2} also requires $L\ll L_{\text{NL}}$ to stay in the no-multi-pair limit. We note that $L\ll L_{\text{D}}$ and $L\ll L_{\text{NL}}$ are just the conditions required for one to be able to neglect dispersion and SPM effects, respectively. Thus one ensures that the pump does not undergo significant dispersion nor SPM as it propagates through a channel waveguide simply by working with pump pulses long enough and weak enough to ensure efficient pair generation per pump pulse with negligible multi-pair generation. More precisely, we find that the no-multi-pair limit for filtered generation \eqref{NC1} in fact demands that the pump power be much less than%
\begin{equation}
P_{\text{f}}^{\text{C}}\equiv \left( \frac{1}{TB}\right) ^{1/2}\left( \gamma
L\right) ^{-1},  \label{PCf}
\end{equation}%
and for the unfiltered \eqref{NC2} case less than%
\begin{equation}
P_{\text{u}}^{\text{C}}\equiv \left( \frac{9\pi L}{2L_{\text{D}}}\right)
^{1/4}\left( \gamma L\right) ^{-1}.  \label{PCu}
\end{equation}%
It will often be possible to refine Eq.~\eqref{PCf} by noting that the filtering is commonly closely matched to the pump bandwidth, so that $TB\sim1$. For example, in the pulsed pump fiber experiment of Liang et al.~\cite{Liang:2007}, $\sqrt{TB}\approx 0.8$.

We can also obtain expressions in the limit of a CW pump ($T\rightarrow\infty $) by performing a Schmidt decomposition on the biphoton wave function:
\begin{equation}
\phi \left( \omega _{1},\omega _{2}\right) =\sum_{\lambda }\sqrt{p_{\lambda }%
}\Phi _{\lambda }\left( \omega _{1}\right) \Phi _{\lambda }\left( \omega
_{2}\right) ,
\end{equation}%
and requiring that the product $\sqrt{p_{1}}\left\vert\beta\right\vert\ll1$, where $\sqrt{p_{1}}$ is largest Schmidt coefficient~\cite{Helt:2012}. The Schmidt value can be approximated by fitting the biphoton wave function to the product of two general Gaussian functions. In this limit, the filtered and unfiltered limits become
\begin{align}
P_{\text{fCW}}^{\text{C}}\equiv & \left( \frac{2\ln \left( 2\right) \pi ^{2}}{64s^{2}}\right) ^{1/4}\left( \gamma L\right) ^{-1}  \notag \\
\approx &\ 0.58\left( \gamma L\right) ^{-1},  \label{PCfCW}
\end{align}%
and%
\begin{align}
P_{\text{uCW}}^{\text{C}}\equiv & \left( \frac{9\pi }{64s}\right)
^{1/4}\left( \gamma L\right) ^{-1}  \notag \\
\approx &\ 0.75\left( \gamma L\right) ^{-1},  \label{PCuCW}
\end{align}%
where $s\approx 1.3916$ is the positive root of $\sinc^{2}\left( x\right)=0.5$. We note that the CW restrictions \eqref{PCfCW} and \eqref{PCuCW} are in fact stronger restrictions than that to neglect SPM. The other two restrictions, \eqref{PCf} and \eqref{PCu}, need not necessarily be stronger, but in many cases they are, as we show below.

\textit{Microring Resonator Expressions. }In contrast to a channel waveguide, for a microring resonator generated photons are typically collected across a single resonance of bandwidth $\Delta _{R}\approx\omega_{P}/Q$ where $Q$ is the quality factor of the ring. Thus, even for modest $Q$ factors, it is not so difficult as in a channel waveguide to generate frequency-uncorrelated photon pairs by using a large enough pump bandwidth~\cite{Helt:2010}. In the short pulse limit $\Delta _{P}\gg \Delta _{R}$, collecting generated photons from a single (paired) resonance on either side of the pumped resonance, and approximating the enhancement factors as Lorentzians, with $k\left(\omega\right)\approx\omega_{P}+\left(\omega-\omega_{P}\right)/v_\text{g}$, we find%
\begin{equation}
N_{\text{pairs}}^{\text{R}}\approx \left( \gamma PL\right) ^{2}\frac{1}{2}%
\left( \frac{Tv_\text{g}}{L}\right) ^{4},  \label{NR2}
\end{equation}%
where $v_\text{g}$ is the group velocity at the pump frequency. However, it is also true that a first experiment in any new system is likely to involve longer pulses or even CW lasers, and so we also consider the long pulse limit $\Delta _{P}\ll \Delta _{R}$. In this limit, we again collect generated photons from a single (paired) resonance on either side of the pumped resonance. Approximating the enhancement factors $F\left(\omega\right)$, rather than the pump pulse wave forms $\phi_{\text{P}}\left(\omega\right)$, as constant in the integral over $\omega $ , we find instead%
\begin{equation}
N_{\text{pairs}}^{\text{R}}\approx \left( \gamma PL\right) ^{2}\frac{v_\text{g}}{2L}%
\left\vert F\left( \omega _{P}\right) \right\vert ^{6}T,  \label{NR3}
\end{equation}%
where the resonant field enhancement factor%
\begin{equation}
F\left(\omega_{P}\right)=2i\sqrt{v_\text{g}Q/\left(\omega_{P}L\right)}.  \label{F}
\end{equation}%
Note that more generally the field enhancement factor for a ring
\begin{equation}
F\left( \omega \right) =\frac{i\kappa }{1-\sigma e^{ik\left( \omega \right)
L}},
\end{equation}%
with $\kappa $ and $\sigma $ the usual cross- and self-coupling coefficients~\cite{Heebner:2008}.

As above, we now consider the no-multi-pair limit and examine the consequences for other nonlinear effects. Recalling \eqref{F}, the restriction $\Delta _{P}\ll \Delta _{R}$ is seen to be equivalent to the condition $T\gg aL\left\vert F\left( \omega _{P}\right)\right\vert ^{2}/v_\text{g}$. Then Eq.~\eqref{NR3} in turn also requires $L\ll L_{\text{NL}}$ to stay within the no-multi-pair limit for generation within a single pair of resonances. On the other hand, for a very short pulse \eqref{NR2}, such a restriction is not required. Again, being a bit more precise, we find that the no-multi-pair limit for a short pulse \eqref{NR2} in fact demands that the pump power be less than%
\begin{equation}
P_{\text{S}}^{\text{R}}\equiv \sqrt{2}\left( \frac{L}{v_\text{g}T}\right) ^{2}\left(
\gamma L\right) ^{-1},  \label{PRS}
\end{equation}%
and for a long pulse%
\begin{equation}
P_{\text{L}}^{\text{R}}\equiv \sqrt{2}\sqrt{\frac{L}{v_\text{g}T}}\frac{1}{\left\vert
F\left( \omega _{P}\right) \right\vert ^{3}}\left( \gamma L\right) ^{-1}.
\label{PRL}
\end{equation}%
In the same CW limit as above, the second of these becomes%
\begin{align}
P_{\text{CW}}^{\text{R}}& \equiv \frac{\left[ \left( \sqrt{2}-1\right)
/\left( 16s^{2}\right) \right] ^{1/4}}{\left\vert F\left( \omega _{P}\right)
\right\vert ^{7/2}}\left( \gamma L\right) ^{-1}  \notag \\
& \approx \frac{0.34}{\left\vert F\left( \omega _{P}\right) \right\vert
^{7/2}}\left( \gamma L\right) ^{-1}.  \label{PRCW}
\end{align}

\textit{Summary. }Reexpressed in terms of power, the condition to neglect XPM is that the pump power be less than%
\begin{equation}
P_{\text{XPM}}\equiv 0.5 \left(\gamma L \right)^{-1}, \label{XPM}
\end{equation}%
with the power that constrains SPM being twice as large. As the pump power near a ring resonance is enhanced by approximately $\left\vert F\left(\omega_P\right)\right\vert^2$, \eqref{XPM} should be divided by this enhancement factor for ring calculations. Thus we see that constraining XPM for pulsed pumps may or may not provide a tighter restriction than that to neglect multi-pair events, depending on filter bandwidths, resonance linewidths, and dispersion. Constraining XPM for CW pumps provides the tightest restriction for channel waveguides, whereas constraining multi-pair events provides the tightest restriction for ring resonators. That is, for CW pumps the no-multi-pair powers \eqref{PCfCW} and \eqref{PCuCW} contain factors in front of $\left( \gamma L\right) ^{-1}$ that are greater than 0.5, \eqref{PRCW} contains a factor less than 0.5, and for pulsed pumps the no-multi pair powers \eqref{PCf}, \eqref{PCu}, \eqref{PRS}, and \eqref{PRL} contain factors in front of $\left( \gamma L\right) ^{-1}$ that depend on the product of pump pulse duration and collection bandwidth, and may be greater than or less than 0.5.

\textit{Absorptive Effects. }We should also consider the absorptive effects of TPA and FCA. We note that there is nothing inherently quantum about their effect on the coherent state pump, and so the constraints from classical nonlinear optics are applicable here. In particular, cross-two-photon absorption of the generated photons can be neglected provided%
\begin{equation}
\frac{\beta _{\text{TPA}}PL}{\mathcal{A}_{\text{eff}}}\ll 1,  \label{TPA}
\end{equation}%
where $\beta _{\text{TPA}}$ is the TPA coefficient \cite{Monat:2009}, or equivalently if the pump power is less than%
\begin{equation}
P_{\text{TPA}}\equiv \frac{1}{2r}\left( \gamma L\right) ^{-1},  \label{TPA2}
\end{equation}%
where $r=\beta _{\text{TPA}}/(2k_{0}n_{2})$ is the nonlinear figure of merit~\cite{Yin:2007}. We note that two-photon absorption of the pump itself requires that the pump power be less than $2P_{\text{TPA}}$. Since for any useful nonlinear material $r<0.5$, this is in fact a weaker requirement than the earlier condition \eqref{XPM}. For materials of interest, the no-XPM limit thus automatically constrains TPA. Similarly, for a pulse, free-carrier absorption can be neglected provided the pump power is less than%
\begin{equation}
P_{\text{FCA}}\equiv \frac{3\hbar \omega _{P}\mathcal{A}_{\text{eff}}}{%
\sigma_\text{FCA} T},  \label{FCA}
\end{equation}%
where $\sigma_\text{FCA} $ is the free-carrier absorption coefficient \cite{Yin:2007}. In the CW limit, free-carrier absorption can be neglected provided the total number of free carriers%
\begin{equation}
n_{\text{tot}}\equiv \frac{n_{\text{SS}}\sigma_\text{FCA} L}{2},
\end{equation}%
is kept much less than unity, where $n_{\text{SS}}$ is the steady-state free carrier density
\begin{equation}
n_{\text{SS}}=\frac{\beta _{\text{TPA}}P^{2}\tau _{c}}{2\hbar \omega _{P}%
\mathcal{A}_{\text{eff}}^{2}},
\end{equation}%
with $\tau _{c}$ the free-carrier lifetime \cite{Yin:2007}. Thus we expect FCA to be negligible in a CW experiment if the pump power is kept less than%
\begin{equation}
P_{\text{CWFCA}}\equiv \left( \frac{4\hbar \omega _{P}\mathcal{A}_{\text{eff}%
}^{2}}{\beta _{\text{TPA}}\tau _{c}\sigma_\text{FCA} L}\right) ^{1/2}. \label{CWFCA}
\end{equation}%
As above, the limiting powers \eqref{TPA2}, \eqref{FCA}, and \eqref{CWFCA} should all be divided by the enhancement factor $\left\vert F\left(\omega_P\right)\right\vert^2$ for ring calculations.

\textit{Conclusions. }Finally, we apply the inequalities developed in this paper to a number of photon pair generation systems. As examples, we consider the photon pair sources in three of the references mentioned above \cite{Liang:2007,Xiong:2011, Azzini:2012}, as well as a potential experiment involving a diamond ring resonator pumped in the short pulse limit, and present the corresponding limits in Table \ref{tab:power}. For all experiments but the CW ring, the no-XPM limit imposes the strongest constraint on the pump power, and thus SPM, multi-pair events, TPA, and FCA can all be safely ignored if one works with pump powers below this limit. For the CW ring, it is the no-multi-pair constraint that is strongest, with all other processes studied here becoming relevant for higher powers. This work suggests that a simple classical calculation can direct designers and users of SFWM photon pair sources to the pump powers at which various nonlinear effects become important.  It is often enough to work to avoid XPM or multi-pair production, and many other nonlinear effects will not be a problem.  In addition to providing such constraints, this work also provides a hierarchy of limiting pump powers that may guide the order in which different effects are incorporated into a more complete treatment.%
\begin{table}[tbp]
\caption{Relevant powers for several real and potential experiments. The second
row is the appropriate pump power to constrain multi-pair production: from
left to right the $P_{\text{multi}}$ used is $P_{\text{f}}^{\text{C}}$, $P_{%
\text{fCW}}^{\text{C}}$, $P_{\text{S}}^{\text{R}}$ and $P_{\text{CW}}^{\text{%
R}}$. For the resonator experiments, the appropriate limiting powers have been divided by the enhancement factor $\left\vert F\left(\omega_P\right)\right\vert^2$. For the rings we have taken $v_\text{g} = c/n_\text{eff}$ with $n_\text{eff}=2.39$ for the diamond ring and $n_\text{eff}=2.47$ for the silicon ring. All values are in Watts.}%
\begin{ruledtabular}
\begin{tabular}{c c c c c}
& Pulsed & CW & Pulsed & CW \\
& Fiber & Waveguide & Ring & Ring \\
& (SiO$_2$)~\cite{Liang:2007} &  (As$_2$S$_3$)~\cite{Xiong:2011} & (D) & (Si)~\cite{Azzini:2012} \\
\hline
$P_{\text{XPM}}$ & 0.77 & 0.50 & 1195 & 0.83 \\
$P_{\text{multi}}$ & 1.96 & 0.58 & 1.1$\times10^7$ & 0.018 \\
$P_{\text{TPA}}$ & $\infty$ & $>$1183 & $\infty$ & 8 \\
$P_{\text{FCA/CWFCA}}$ & $\infty$ & $\infty$ & $\infty$ & 0.06 \\
\end{tabular}
\end{ruledtabular}
\label{tab:power}
\end{table}

\begin{acknowledgments}
This work was supported in part by the ARC Centre for Ultrahigh bandwidth Devices for Optical Systems (CUDOS) (project number CE110001018), and the Natural Sciences and Engineering Research Council of Canada (NSERC).
\end{acknowledgments}

\bibliography{kickin}

\end{document}